\documentclass[11pt]{article}
\usepackage[margin=1in]{geometry}
\usepackage{amsmath,graphicx}
\usepackage{amsfonts}
\usepackage{caption}
\usepackage{hyperref}
\usepackage{subcaption}
\usepackage{algorithm}
\usepackage{algorithmic}
\usepackage{mathtools} 
\usepackage{cite}
\usepackage{multirow}
\usepackage{tikz}

\def\mbf{\mathbf}

\DeclareMathOperator{\argmin}{\mathrm{argmin}}

\newcommand{\comment}[1]{}
\usepackage{xcolor}

\usepackage[normalem]{ulem}
\newcommand\redout{\bgroup\markoverwith{\textcolor{red}{\rule[.5ex]{2pt}{0.4pt}}}\ULon}

\usepackage{lipsum}

\newcommand\blfootnote[1]{%
  \begingroup
  \renewcommand\thefootnote{}\footnote{#1}%
  \addtocounter{footnote}{-1}%
  \endgroup
}

\begin{document}
%
\title{Low-Rank Phase Retrieval with Structured Tensor Models\blfootnote{The work of the authors was supported by the US NSF under award CCF-1910110. The authors are with the Department of Electrical and Computer Engineering, Rutgers, The State University of New Jersey, New Brunswick, NJ 08854, \texttt{smk330@scarletmail.rutgers.edu}, \texttt{xl598@soe.rutgers.edu}, \texttt{ads221@soe.rutgers.edu}. Code and supplementary results are available at \href{https://gitlab.com/sarwate_lab/Tucker-Structured-Phase-Retrieval/-/tree/main}{\texttt{https://gitlab.com/sarwate\_lab}}.}}
%
%
%

\author{Soo Min Kwon,
        Xin Li,
        Anand D. Sarwate}
        
\date{\today}

%
%

\markboth{Journal of \LaTeX\ Class Files,~Vol.~14, No.~8, August~2015}%
{Shell \MakeLowercase{\textit{et al.}}: Bare Demo of IEEEtran.cls for IEEE Journals}
%



\maketitle

\begin{abstract}
We study the low-rank phase retrieval problem, where the objective is to recover a sequence of signals (typically images) given the magnitude of linear measurements of those signals.
Existing solutions involve recovering a matrix constructed by vectorizing and stacking each image. These algorithms model this matrix to be low-rank and leverage the low-rank property to decrease the sample complexity required for accurate recovery. 
However, when the number of available measurements is more limited, these low-rank matrix models can often fail. 
We propose an algorithm called Tucker-Structured Phase Retrieval (TSPR) that models the sequence of images as a tensor rather than a matrix that we factorize using the Tucker decomposition. This factorization reduces the number of parameters that need to be estimated, allowing for a more accurate reconstruction in the under-sampled regime. Interestingly, we observe that this structure also has improved performance in the over-determined setting when the Tucker ranks are chosen appropriately. 
We demonstrate the effectiveness of our approach on real video datasets under several different measurement models.

\end{abstract}


%

\section{Introduction}
\label{sec:intro}
Phase retrieval, or quadratic sensing, is a problem that arises from a wide range of imaging domains such as X-ray crystallography~\cite{crystal}, Fourier ptychography~\cite{fourier_one, fourier_two}, and astronomy~\cite{astronomy}. In each of these domains, the measurement acquisition process generally involves an optical sensor that captures the diffracted patterns of the object of interest. However, the physical limitations of these sensors only allow us to observe the intensities (or magnitudes) of these patterns. The objective of phase retrieval is then to recover this object $\mbf{x} \in \mathbb{C}^n$, given a sampling matrix $\mbf{A} \in \mathbb{C}^{n \times m}$ and measurements $\mbf{y} \in \mathbb{R}^m$, where
\begin{align}
    \mbf{y} = |\mbf{A}^{*} \mbf{x}|,    
\end{align}
(or equivalently, $\mbf{y} = |\mbf{A}^{*} \mbf{x}|^2$) where $*$ represents the Hermitian (or conjugate) transpose.
The importance of solving the phase retrieval problem in these imaging domains have led to many convex and non-convex solutions~\cite{phase_matrix_completion, phase_lift, altminphase, phase_wf, twf, rwf_two}. However, the theoretical guarantees of all existing methods require the system to be over-determined (i.e. $m \gg n$). This requirement, which is considered to be the bottleneck of phase retrieval, mainly comes from the non-convex nature of the problem. In order to converge to the optimal solution, one needs enough samples to guarantee that the initial estimate of the signal is close to the true signal with high probability. This initial estimation step is called spectral initialization, where the term ``spectral'' comes from the use eigenvectors (or singular vectors) of properly designed matrices from data~\cite{chen2021spectral}. This step has been shown to be essential for solving the phase retrieval problem, and many variants of this step have been proposed in the literature.


Recently, there has been a surge of interest in solving the \emph{low-rank phase retrieval} problem~\cite{lrpr, prov_lrpr, lrpr_fourier, altmintrunc, bayesian_lrpr}. This problem can be viewed as a dynamic extension of the standard phase retrieval problem, where the objective is to recover a matrix of vectorized images rather than a single image. Formally, we want to estimate a low-rank matrix $\mbf{X} \in \mathbb{C}^{n \times q}$, where
\begin{align}
    \mbf{X} = [\mbf{x}_1, \mbf{x}_2 \ldots, \mbf{x}_q],
\end{align}
with $\mbf{x}_k \in \mathbb{C}^n$ given sampling matrices $\mbf{A}_k \in \mathbb{C}^{n\times m}$ and measurements 
\begin{align}
    \mbf{y}_k = |\mbf{A}_k^{*} \mbf{x}_k|, \, k=1,\ldots, q.
\end{align}
In this problem formulation, we assume that there is a separate, independent set of sampling matrices $\mbf{A}_k$ for each signal $\mbf{x}_k$. Unlike the phase retrieval problem, this problem has several solutions that have strong theoretical guarantees even for the under-determined setting (i.e. $m \ll n$). These algorithms exploit the low-rank property of the matrix $\mbf{X}$ with the extra set of sampling matrices in order to naturally reduce the sample complexity. 
However, our empirical results suggest that there is perhaps a gap between theory and practice, and that these solutions fail to accurately recover the images in the under-determined setting.
In fact, in these settings, we observe that these algorithms often do not converge.

In this paper, we propose an algorithm called Tucker-Structured Phase Retrieval (TSPR) that models the sequence of images as a tensor rather than a matrix. With a tensor model, we can decompose the tensor using the Tucker decomposition~\cite{Kolda} to estimate fewer parameters than the matrix counterpart. The reduction in the number of parameters also decreases the number of degrees of freedom, suggesting that the recovery of the sequence of signals is possible with a smaller sample complexity. In the literature, it has been shown that this idea of modelling the parameters as a tensor have been effective in solving many other statistical estimation problems~\cite{GhassemiSSB:20separable, li2018tucker, zhang2020islet}.
We adopt the idea for low-rank phase retrieval and empirically show that recovery is indeed possible with a smaller number of measurements. We conduct experiments on real video datasets with measurements generated from real and complex Gaussian vectors and coded diffraction patterns. Our results show that in all of these measurement settings, our algorithm outperforms existing algorithms in both the under and over-determined regimes.

\noindent\textbf{Notation:} We denote scalars with lowercase letters (e.g. $x$), vectors with bold lowercase letters (e.g. $\mbf{x}$), matrices with bold uppercase letters (e.g. $\mbf{X}$), and tensors with underlined, bold uppercase letters (e.g. $\underline{\mbf{X}}$). We denote the $n$-th column of the matrix $\mbf{X}$ as $\mbf{x}_n$. Similarly, we denote the $n$-th frontal slice of the tensor $\underline{\mbf{X}}$ as $\mbf{X}_n$. Lastly, we denote the inner product between two vectors $\mbf{a}$ and $\mbf{b}$ as $\langle \mbf{a}, \mbf{b} \rangle$.

\section{Unstructured Low-Rank Phase Retrieval}

There are several provably efficient algorithms for solving the low-rank phase retrieval problem that vectorize each image and recover a low-rank matrix. We call such methods ``unstructured'' because they assume no structure in the images. Recently, Nayer et al.~proposed AltMinLowRaP~\cite{prov_lrpr}, an algorithm that theoretically improved their previous algorithm AltMinTrunc~\cite{altmintrunc, lrpr}, that both solved the unstructured low-rank phase retrieval problem. AltMinLowRaP involved alternately minimizing the factor matrices $\mbf{U} \in \mathbb{C}^{n \times r}$ and $\mbf{B} \in \mathbb{C}^{q \times r}$ that constructed the low-rank matrix $\mbf{X} = \mbf{U} \mbf{B}^{*}$. Updating the factor matrix $\mbf{U}$ consisted of minimizing the objective function
\begin{align}
    \underset{\mbf{U}}{\argmin} \, \sum_k  \left\lVert \mbf{C}_k \mbf{y}_k - \mbf{A}_k^* \mbf{U}\mbf{b}_k \right\rVert_2^2,
\label{eq:update_u}
\end{align}
where $\mbf{b}_k$ is the $k$-th row of the matrix $\mbf{B}$ and $\mbf{C}_k$ is a diagonal phase matrix. Note that this objective function sums over all of the columns in $\mbf{X}$, as the $k$-th column of $\mbf{X}$ can be written as $\mbf{x}_k = \mbf{U}\mbf{b}_k$. The intuition behind this summation can be viewed as each of the vectorized images $\mbf{x}_k$ differing by $\mbf{b}_k$, while sharing the same $\text{span}(\mbf{U})$. Optimizing for $\mbf{U}$ involved minimizing this objective function using conjugate gradient least squares (CGLS) while keeping $\mbf{b}_k$ fixed. The factor matrix $\mbf{B}$ was initialized and updated by solving an $r$-dimensional noisy phase retrieval problem for each row of $\mbf{B}$, $\mbf{b}_k$. To see this, we can rewrite each of the measurements as
\begin{align}
    y_{i, k} &= |\langle \mbf{a}_{i, k}, \mbf{x}_k \rangle| \\
    &= |\langle \mbf{a}_{i, k}, \mbf{U}\mbf{b}_k \rangle| = |\langle \mbf{U}^{*} \mbf{a}_{i, k}, \mbf{b}_k \rangle|.
\end{align}
Given an estimate of $\mbf{U}$, we can solve for each $\mbf{b}_k$ using any phase retrieval method, such as Reshaped Wirtinger Flow (RWF)~\cite{rwf_two}. Thus, AltMinLowRaP runs RWF $q$ times (once for each image) to estimate $\mbf{b}_k$ given the sampling matrix $\mbf{U}^{*} \mbf{a}_{i, k}$. Lastly, upon updating the matrix $\mbf{U}$ and each vector $\mbf{b}_k$, the phase matrices were also updated by taking the phases of $\mbf{x}_k = \mbf{U}\mbf{b}_k$ as follows:
\begin{align}
    \mbf{C}_k = \text{Diag}(\text{Phase}(\mbf{A}_k^* \mbf{U}\mbf{b}_k)).
\end{align}

Due to the non-convex nature of this problem, the factor matrix $\mbf{U}$ was also initialized via a spectral method. The matrix $\mbf{U}$ was initialized by taking the top $r$ eigenvectors of the surrogate matrix
\begin{align}
    \mbf{Y} = \frac{1}{mq} \sum_{i=1}^m \sum_{k=1}^q y_{i, k}^2 \mbf{a}_{i, k} \mbf{a}_{i, k}^{*} \mbf{1}_{\{y_{i, k}^2 \leq \frac{\alpha^2}{mq} \sum_{t, v} y_{t,v}^2\}},
\end{align}
for some trimming threshold $\alpha$. 
The intuition behind this matrix is that given enough samples, the expectation of this matrix is equivalent to
\begin{align}
    \mathbb{E}[y_{i, k} \mbf{a}_{i, k} \mbf{a}_{i, k}^{*}] = 2\mbf{x}_k\mbf{x}_k^{*} + \lVert \mbf{x}_k \rVert^2 \mbf{I}.
\end{align}
Thus, the subspace spanned by the top $r$ eigenvectors of $\mbf{Y}$ can recover exactly $\mbf{U}$. The double summation over the measurements and samples in the surrogate matrix and truncation is what guaranteed AltMinLowRaP a smaller sample complexity over existing methods. Our algorithm is an improvement over AltMinLowRaP that empirically works better in both the under and (some) over-sampled regimes. Although our algorithm does not yet have a theoretical analysis of the sample complexity, our results show that our algorithm can work better in practice.

\section{Tucker-Structured Phase Retrieval}
\label{sec:algorithm}

Our algorithm models the sequence of $q$ images as a tensor by reshaping and stacking each of the vectorized images from $\mbf{x}_k \in \mathbb{C}^n$ into $\mbf{X}_k \in \mathbb{C}^{n_1 \times n_2}$, where $n = n_1 n_2$.
The objective of TSPR is to recover this tensor $\underline{\mbf{X}} \in \mathbb{C}^{n_1 \times n_2 \times q}$, where $\underline{\mbf{X}}$ can be factorized using the Tucker decomposition written as 
\begin{align}
    \underline{\mbf{X}} = \underline{\mbf{G}} \times_1 \mbf{D} \times_2 \mbf{E} \times_3 \mbf{F},
\end{align}
where $\underline{\mbf{G}} \in \mathbb{C}^{r_1 \times r_2 \times r_3}$ is the core tensor and  $\mbf{D} \in \mathbb{C}^{n_1 \times r_1}$, $\mbf{E} \in \mathbb{C}^{n_2 \times r_2}$, and $\mbf{F} \in \mathbb{C}^{q \times r_3}$ are the factor matrices. The values $r_1, r_2,$ and $r_3$ correspond to the ranks of each dimension of the tensor. Specifically, $r_1$ and $r_2$ refer to the ranks of the frontal slices of the tensor (an image), whereas $r_3$ refers to the temporal rank that corresponds to the “rank” in the standard model which vectorizes the images.  We want to solve for these factors by first initializing them via a spectral method and then estimating them using alternating minimization and CGLS.

\noindent \textbf{Spectral Initialization:} The idea behind our spectral initialization step is to construct a tensor that is close to $\underline{\mbf{X}}$ with high probability. Once we construct this tensor, we can use higher-order SVD (HOSVD)~\cite{hosvd} to initialize our core tensor and factor matrices.
We adopt the initialization technique of Truncated Wirtinger Flow (TWF)~\cite{twf} to obtain an initial estimate of the vectorized image $\mbf{x}_k$. Specifically, we want to first take the leading eigenvector of the constructed matrix
\begin{align}
    \mbf{Y}_k = \sum_{i=1}^m y_{i, k}^2 \mbf{a}_{i, k} \mbf{a}_{i, k}^{*} \mbf{1}_{ \{|y_{i,k}|^2 \leq \alpha^2 \lambda_k^2 \} },
\end{align}
where 
\begin{align}
    \lambda_k = \sqrt{\frac{1}{m}\sum_{i=1}^m y_{i, k}}.
\end{align} If $\mbf{z}_k$ is the leading eigenvector of $\mbf{Y}_k$, we compute the initial estimate of $\mbf{x}_k$ as 
\begin{align}
\mbf{x}_k = \sqrt{\frac{mn}{\sum_{i=1}^m \lVert \mbf{a}_{i, k}\rVert_2^2}} \lambda_k \mbf{z}_k,    
\end{align}
which appropriately normalizes $\mbf{z}_k$ to approximately have the same norm as $\mbf{x}_k$. Upon computing each $\mbf{x}_k$ for $k=1, \ldots, q$, we reshape $\mbf{x}_k$ back into its original dimensions and stack them to create the initial tensor. This initialization step is outlined in Algorithm~\ref{alg:tspr_init}.

\noindent \textbf{Alternating Minimization:} Upon initialization, we can alternately update the core tensor and each factor matrix using CGLS and RWF. Recall that in AltMinLowRaP, we minimized an objective function that was formed by plugging in $\mbf{x}_k = \mbf{U}\mbf{b}_k$. Similarly, we can minimize the same function, but by rewriting $\mbf{x}_k$ using our Tucker factors. In specific, we can write each $\mbf{x}_k$ as 
\begin{align}
    \mbf{x}_k = (\mbf{f}_k \otimes \mbf{E} \otimes \mbf{D})\text{vec}(\underline{\mbf{G}}),
\end{align}
where $\mbf{f}_k$ is the $k$-th row of the factor matrix $\mbf{F}$. The reason behind writing $\mbf{x}_k$ in terms of $\mbf{f}_k$ is the same reasoning used for the unstructured case -- each image $\mbf{x}_k$ differs by $\mbf{f}_k$.
By plugging in $\mbf{x}_k$, the update steps of the core tensor $\underline{\mbf{G}}$ and factor matrices $\mbf{D}$ and $\mbf{E}$ consists of minimizing the function
\begin{align}
    \sum_k  \left\lVert \mbf{C}_k \mbf{y}_k - \mbf{A}_k^* (\mbf{f}_k \otimes \mbf{E} \otimes \mbf{D})\text{vec}(\underline{\mbf{G}}) \right\rVert_2^2.
\label{eq:tensor_x_k}
\end{align}
To update each row vector $\mbf{f}_k$, note that we can rewrite $y_{i, k}$ as 
\begin{align}
y_{i, k} &= |\langle \mbf{a}_{i, k}, \mbf{x}_k \rangle|\\
&= |\langle \mbf{a}_{i, k}, \mathcal{M}_3(\underline{\mbf{G}})(\mbf{E} \otimes \mbf{D})^{*} \mbf{f}_k \rangle| = |\langle \mathcal{M}_3(\underline{\mbf{G}})(\mbf{E} \otimes \mbf{D})^{*} \mbf{a}_{i, k}, \mbf{f}_k \rangle|,
\end{align}
where $\mathcal{M}_k(\underline{\mbf{G}})$ is the $k$-th mode matricization of the tensor $\underline{\mbf{G}}$.
With this formulation, updating each $\mbf{f}_k$ simplifies to solving a noisy $r$-dimensional phase retrieval problem with sampling matrix $\mathcal{M}_3(\underline{\mbf{G}})(\mbf{E} \otimes \mbf{D})^{*} \mbf{a}_{i, k}$. We can use any classical phase retrieval method to solve for $\mbf{f}_k$, but we use 
RWF~\cite{rwf_two} to directly compare to AltMinLowRaP. This update step is summarized in Algorithm~\ref{alg:tspr}, and the details for implementation are available in the Appendix.

\begin{algorithm}[h!]
\caption{TSPR Initialization}
\begin{algorithmic}[1]
\REQUIRE Observations: $\{y_{i, k} \,|\, 1 \leq i \leq m, 1 \leq k \leq q \}$, Sampling vectors: $\{ \mbf{a}_{i, k} \,|\, 1 \leq i \leq m, 1 \leq k \leq q\}$, Trimming threshold: $\alpha$, ranks $=[r_1, r_2, r_3]$
\FOR {$k = 1, \hdots, q$}
\STATE Compute $\lambda_k = \sqrt{\frac{1}{m}\sum_{i=1}^m y_{i, k}}$.
\STATE Compute $\mbf{z}_k$ as leading eigenvector of\;
    \begin{align*}
        \mbf{Y}_k = \sum_{i=1}^m y_{i, k}^2 \mbf{a}_{i, k} \mbf{a}_{i, k}^{*} \mbf{1}_{ \{|y_{i,k}|^2 \leq \alpha^2 \lambda_k^2 \} }
    \end{align*}
\STATE Compute $\mbf{x}_k = \sqrt{\frac{mn}{\sum_{i=1}^m \lVert \mbf{a}_{i, k}\rVert_2^2}} \lambda_k \mbf{z}_k$.
\STATE Reshape $\mbf{x}_k \in \mathbb{C}^{n}$ into $\mbf{X}_k \in \mathbb{C}^{n_1 \times n_2}$.
\ENDFOR
\STATE Stack tensor into $\underline{\mbf{X}} = [\mbf{X}_1, \mbf{X}_2, \ldots, \mbf{X}_q]$
\STATE Initialize factors using HOSVD:\;
\begin{align*}
    \mbf{D}^0, \mbf{E}^0, \mbf{F}^0, \mbf{\underline{\mbf{G}}}^0 = \text{HOSVD}(\underline{\mbf{X}}, \text{ranks}) 
\end{align*}
\ENSURE $\mbf{D}^0, \mbf{E}^0, \mbf{F}^0, \mbf{\underline{\mbf{G}}}^0$
\end{algorithmic}
\label{alg:tspr_init}
\end{algorithm}

\begin{algorithm}[h!]
\caption{Tucker-Structured Phase Retrieval (TSPR)}
\begin{algorithmic}[1]
\REQUIRE Observations: $\{y_{i, k} \,|\, 1 \leq i \leq m, 1 \leq k \leq q \}$, Sampling vectors: $\{ \mbf{a}_{i, k} \,|\, 1 \leq i \leq m, 1 \leq k \leq q\}$, Initial factors: $\mbf{D}^0, \mbf{E}^0, \mbf{F}^0, \underline{\mbf{G}}^0$, Iterations $T$, RWF Iterations $T_{RWF}$
\FOR {$t = 1, \ldots, T$}
\FOR{$k= 1, \ldots, q$}
    \STATE Update $\mbf{f}_k^{t+1} = \text{RWF}([\mbf{D}^t, \mbf{E}^t, \underline{\mbf{G}^t}, \mbf{A}_k^{*}], \mbf{y}_k, T_{RWF})$
    \STATE Compute $\mbf{X}_k^{t+1} = (\mbf{f}_k^{t+1} \otimes \mbf{E}^t \otimes \mbf{D}^t)\text{vec}(\underline{\mbf{G}}^t)$
    \STATE Update diagonal phase matrix $\mbf{C}_k^{t+1} = \text{Diag}(\text{Phase}(\mbf{A}_k^{*} \text{vec}(\mbf{X}_k^{t+1})))$\ENDFOR
    \STATE Update $\mbf{D}^{t+1}$, $\mbf{E}^{t+1}$, $\underline{\mbf{G}}^{t+1}$ by minimizing (\ref{eq:tensor_x_k})
\ENDFOR
\STATE Reconstruct tensor $\underline{\mbf{X}}^T = \underline{\mbf{G}}^T \times_1 \mbf{D}^T \times_2 \mbf{E}^T \times_3 \mbf{F}^T$
\ENSURE $\underline{\mbf{X}}^T$
\end{algorithmic}
\label{alg:tspr}
\end{algorithm}

\section{Numerical Experiments}

We compare the performance of TSPR with two closely related algorithms, AltMinTrunc and AltMinLowRaP, using two real video datasets, Mouse and Plane. We consider measurements generated by real Gaussian matrices, complex Gaussian matrices, and coded diffraction patterns (CDP). To quantitatively compare these algorithms, we use the phase-invariant matrix distance~\cite{prov_lrpr} defined as
\begin{align}
    \text{mat-dist}^2(\mbf{\hat{X}}, \mbf{X}) = \sum_{k=1}^q \text{dist}^2(\hat{\mbf{x}}_k, \mbf{x}_k),
\label{eq:mat_dist}
\end{align}
where $\mbf{X}$ is the true matrix, $\hat{\mbf{X}}$ is the reconstructed matrix and
\begin{align}
    \text{dist}(\mbf{\hat{x}}, \mbf{x}) = \underset{\phi \in [0, 2\pi]}{\min} \lVert \mbf{x} - e^{\sqrt{-1}\phi}\hat{\mbf{x}} \rVert.
\end{align}
Note that the distance metric above is written in terms of the columns of the matrices $\mbf{X}$ and $\mbf{\hat{X}}$.
Some of the results went through a ``model correction'' step as proposed by Nayer et al.~\cite{prov_lrpr}.  We provide additional information on this correction step in the Appendix. We also provide a reconstruction of the videos as a supplement and display single frames in this paper.

\begin{table*}[b!]
\centering
\begin{tabular}{ |c|c|c|c|c|c|c| }
    \hline
    Experiment & Samples & \# of Parameters & Algorithm & Rank & Distance \\ \hline
    Mouse (Real) & $m=0.25n$ & $5750$ & TSPR & $r=[20,25, 5]$ & $2.851$ \\ 
         &  & $16450$ & AltMinLowRaP & $r=5$ & $6.175$ \\ 
        & & $16450$ & AltMinTrunc & $r=5$ & 7.277 \\ \hline
    Mouse (Complex) & $m=0.75n$ & $5750$ & TSPR & $r=[20,25, 5]$ & $1.217$ \\ 
        &  & $8700$ &  & $r=[20,25, 10]$ & $1.170$\\ 
         & & $16450$ & AltMinLowRaP & $r=5$ & $4.379$ \\ 
        &  & $32900$& & $r=10$ & $3.435$ \\ \
         &  & $16450$& AltMinTrunc & $r=5$ & $78.118$ \\ 
        &  & $32900$ & & $r=10$ & $77.319$ \\ \hline
    Plane (CDP) & $m=2n$ & $5600$ & TSPR & $r=[15,20, 10]$ & $0.437$ \\ 
     &  & $8075$ & & $r=[20,25, 10]$ & $0.571$\\ 
    &  &  $14525$ & & $r=[30,35, 10]$ & $1.008$ \\ 
         &  & $22900$ & AltMinLowRaP & $r=10$ & $0.869$\\ 
        & & $22900$& AltMinTrunc & $r=10$ & $0.894$ \\ \hline
\end{tabular}
\caption{Results for the experiments with the Mouse and Plane datasets. The value $n$ refers to the dimensions of $\mbf{x}_k$ and $m$ refers to the number of measurements generated for each $\mbf{x}_k$. The \# of parameters value refers to the total number of parameters that need to be solved for all images $\mbf{x}_k$. The distance metric is the phase-invariant distance defined in equation (\ref{eq:mat_dist}).}
\label{table:normalized_errors}
\end{table*}

\begin{figure}[t!]
\begin{subfigure}{.245\textwidth}
  \centering
  \includegraphics[width=\textwidth]{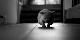} 
\end{subfigure}
\begin{subfigure}{.245\textwidth}
  \centering
  \includegraphics[width=\textwidth]{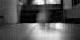}
\end{subfigure}
\begin{subfigure}{.245\textwidth}
  \centering
  \includegraphics[width=\textwidth]{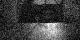} 
\end{subfigure}
\begin{subfigure}{.245\textwidth}
  \centering
  \includegraphics[width=\textwidth]{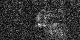} 
\end{subfigure}
\newline
\begin{subfigure}{.245\textwidth}
  \centering
  \includegraphics[width=\textwidth]{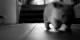}
  \caption*{Original}
\end{subfigure}
\begin{subfigure}{.245\textwidth}
  \centering
  \includegraphics[width=\textwidth]{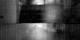} 
  \caption*{TSPR}
\end{subfigure}
\begin{subfigure}{.245\textwidth}
  \centering
  \includegraphics[width=\textwidth]{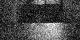} 
  \caption*{AltMinLowRaP}
\end{subfigure}
\begin{subfigure}{.245\textwidth}
  \centering
  \includegraphics[width=\textwidth]{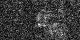} 
  \caption*{AltMinTrunc}
\end{subfigure}
\caption{Results from recovering a video of a moving mouse from complex Gaussian measurements. Rows 1 and 2: reconstructed images of frames 60 and 70, respectively.}
\label{fig:mouse_experiment}
\end{figure}

\noindent \textbf{Experiments with the Mouse Dataset:} 
The mouse dataset is a video of a mouse moving slowly towards a camera, provided by Nayer et al.~\cite{prov_lrpr}. The mouse video consisted of $90$ frames, where each frame was downsized to be of dimensions $40 \times 80$. Upon constructing the tensor $\underline{\mbf{X}} \in \mathbb{C}^{40 \times 80 \times 90}$, we generated measurements according to the model
\begin{align}
    \mbf{y}_k = |\mbf{A}_k^{*} \text{vec}(\mbf{X}_k)|, \, k=1,\ldots, q,
\end{align}
where each column of $\mbf{A}_k$ was drawn either from $\mbf{a}_{i,k} \sim \mathcal{N}(0, \mbf{I})$ (real Gaussian distribution) or $\mbf{a}_{i,k} \sim \mathcal{C}\mathcal{N}(0, \mbf{I})$ (circularly complex Gaussian distribution). We compare the three algorithms in two under-determined settings under these measurements. The numerical results are recorded in Table \ref{table:normalized_errors} with two of the reconstructed frames shown in Figure~\ref{fig:mouse_experiment}. In Table \ref{table:normalized_errors}, we can see that TSPR outperformed the other two algorithms in both under-determined settings by estimating significantly less parameters. In fact, we observe that for two different ranks, AltMinTrunc did not converge and had a resulting error that was significantly higher than the others. These values were obtained by running $T=20$ iterations of the total algorithm and $T_{RWF}=25$ where applicable. We would like to note that each iteration of TSPR also runs several iterations of CGLS. For our experiments, we ran $T_{CGLS} = 50$ iterations, which results in a total of $1000$ iterations, excluding the iterations from RWF. For the trimming threshold, we used a value of $\alpha =3$, as suggested in TWF~\cite{twf}. The ranks were generally chosen by trial and error, and the results did not go through a model correction step, as it seemed to increase the errors both numerically and visually. We would also like to note that even though TSPR yielded a lower numerical reconstruction error, we can see in Figure~\ref{fig:mouse_experiment} that the reconstructed image is still not as clear as the original image.
This is an intrinsic tradeoff of the Tucker model, as each frame may not be exactly low-rank. We want to choose the ranks corresponding to the image dimensions (i.e. $r_1, r_2$) to be small so that we can get convergence up to some modelling error, but not too small such that the reconstructed images are unclear. Based on our experiments, we observed that for ranks $r_1$ and $r_2$, using ranks slightly less than half of the dimensions of image (i.e. $r_1 < 0.5n_1$ and $r_2 < 0.5n_2$) worked well, whereas for $r_3$ (or $r$ in the matrix model), we can be more conservative in our choices and choose a value much smaller.

\begin{figure}[t!]
     \centering
     \begin{subfigure}[b]{0.24\textwidth}
         \centering
         \includegraphics[width=\textwidth]{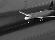}
     \end{subfigure}
     \hfill
     \begin{subfigure}[b]{0.24\textwidth}
         \centering
         \includegraphics[width=\textwidth]{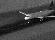}
     \end{subfigure}
     \hfill
     \begin{subfigure}[b]{0.24\textwidth}
         \centering
         \includegraphics[width=\textwidth]{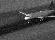}
     \end{subfigure}
     \hfill
         \begin{subfigure}[b]{0.24\textwidth}
         \centering
         \includegraphics[width=\textwidth]{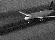}
     \end{subfigure}
     \newline
          \begin{subfigure}[b]{0.24\textwidth}
         \centering
         \includegraphics[width=\textwidth]{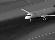}
         \caption*{Original}
     \end{subfigure}
     \hfill
     \begin{subfigure}[b]{0.24\textwidth}
         \centering
         \includegraphics[width=\textwidth]{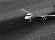}
         \caption*{TSPR}
     \end{subfigure}
     \hfill
     \begin{subfigure}[b]{0.24\textwidth}
         \centering
         \includegraphics[width=\textwidth]{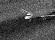}
         \caption*{AltMinLowRaP}
     \end{subfigure}
     \hfill
         \begin{subfigure}[b]{0.24\textwidth}
         \centering
         \includegraphics[width=\textwidth]{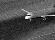}
         \caption*{AltMinTrunc}
     \end{subfigure}
        \centering
        \caption{Results from recovering a video of a plane from CDP measurements. Rows 1 and 2: reconstructed images of frames 10 and 80, respectively.}
        \label{fig:plane_experiment}
\end{figure}

\noindent \textbf{Experiments with the Plane Dataset:} The plane dataset is a video of a plane slowly landing on a runway, also provided by Nayer et al.~\cite{prov_lrpr}. The plane video consisted of $90$ frames, where each frame was downsized to be of dimensions $40 \times 55$ for efficiency. 
Upon constructing the tensor
$\underline{\mbf{X}} \in \mathbb{C}^{40 \times 55 \times 90}$, we generated measurements according to the CDP model
\begin{align}
    \mbf{y}_{l, k} = |\tilde{\mbf{F}} \mbf{M}_l \text{vec}(\mathbf{X}_k)|, \, l=1,\ldots, L, \, k=1,\ldots q,
\end{align}
where $\tilde{\mbf{F}}$ is the discrete Fourier transform (DFT) matrix and $\mbf{M}$ is a diagonal mask matrix with elements drawn randomly from $\{1, -1, j, -j\}$ (details provided in the full version).
Since the CDP model can only generate measurements $m = Ln$ for each image for some integer $L$, the objective of this experiment was to show the effectiveness of TSPR in the over-determined setting. Upon running all three algorithms with the same parameters as the Mouse dataset, each result went through a model correction step. In Figure~\ref{fig:plane_experiment}, we see that while all three algorithms can visually reconstruct the frames of this video, but Table \ref{table:normalized_errors} shows that the error for TSPR is significantly lower. However, the errors are only lower for certain values of the Tucker rank. This is most likely because as these ranks increase, the total number of parameters slowly converge to that of the unstructured methods, making recovery much more difficult.

\section{Conclusion}
In this paper, we showed that by modeling the sequence of images as a tensor, we can obtain 
a more accurate reconstruction in both the under and over-sampled regimes. 
Our algorithm, TSPR, adopted a mixture of optimization techniques from AltMinLowRaP and Truncated Wirtinger Flow to improve upon existing methods. TSPR involved a spectral initialization method that used higher-order SVD with alternating minimization via conjugate gradient least squares. Currently, TSPR lacks the theoretical guarantees in comparison to unstructured solutions. One important avenue for future research can be to extend our algorithm but with theoretical guarantees on the sample complexity required for accurate recovery. Our results show that there \emph{exist} Tucker-structured models with better performance; we believe that perhaps finding a more principled approach for choosing these ranks is an important challenge for future work.

\clearpage

\appendix
\section{Factor Updates with CGLS}
Tucker-Structured Phase Retrieval (TSPR) uses conjugate gradient least squares (CGLS) to update the Tucker factors and core tensor.
In order to use CGLS,
we need to rewrite the objective function in terms of the vectorized factors. Recall that when solving for $\mbf{x}_k$, the objective function that we want to minimize is
\begin{align}
     \underset{\mbf{x}_k}{\argmin} \, \sum_k  \left\lVert \mbf{C}_k \mbf{y}_k - \mbf{A}_k^* \mbf{x}_k \right\rVert_2^2.
\end{align}
In order to update the matrix $\mbf{D}$, we need to rewrite $\mbf{x}_k$ in terms of $\text{vec}(\mbf{D})$ as follows:
\begin{align}
    \mbf{x}_k = \text{vec}(\mbf{D} \cdot \mathcal{M}_1(\underline{\mbf{G}})(\mbf{f}_k \otimes \mbf{E})^{*}).
\end{align}
If we let $\mbf{S}_k = \mathcal{M}_1(\underline{\mbf{G}})(\mbf{f}_k \otimes \mbf{E})^{*}$,
then
\begin{align}
    \mbf{x}_k &= \text{vec}(\mbf{D} \mbf{S}_k) \\
    &= \text{vec}(\mbf{I} \mbf{D} \mbf{S}_k) \\
    &= (\mbf{S}_k^{*} \otimes \mbf{I}) \text{vec}(\mbf{D}),
\end{align}
where $\mbf{I}$ is the identity matrix and the last equality comes from using the property
\begin{align}
    \text{vec}(\mbf{A} \mbf{X} \mbf{B}) = (\mbf{B}^{*} \otimes \mbf{A}) \text{vec}(\mbf{X}),
\end{align}
for any arbitrary matrices $\mbf{A}, \mbf{X}$, and $\mbf{B}$. Thus, by rewriting the objective function as 
\begin{align}
    \sum_k \left\lVert \mbf{C}_k \mbf{y}_k -\mbf{A}_k^{*}  (\mbf{S}_k^{*} \otimes \mbf{I}) \text{vec}(\mbf{D})) \right\rVert^2,
\end{align}
we can solve for $\text{vec}(\mbf{D})$ using CGLS. Similarly, to update factor matrix $\mbf{E}$, we can write $\mbf{x}_k$ as
\begin{align}
    \mbf{x}_k = \text{vec}(\mbf{E} \cdot \mathcal{M}_2(\underline{\mbf{G}})(\mbf{f}_k \otimes \mbf{D})^{*})^{*}.
\end{align}
Let $\mbf{U}_k = \mathcal{M}_2(\underline{\mbf{G}})(\mbf{f}_k \otimes \mbf{D})^{*}$. Then,
\begin{align}
    \mbf{x}_k &= \text{vec}((\mbf{I}\mbf{E} \mbf{U}_k)^{*} ) \\
    &= \text{vec}(\mbf{U}_k^{*}\mbf{E}^{*} \mbf{I}^{*}) \\
    &= (\mbf{I} \otimes \mbf{U}_k^{*}) \text{vec}(\mbf{E}^{*}).
\end{align}
The update step for $\mbf{E}$ becomes minimizing the objective function
\begin{align}
    \sum_k \left\lVert \mbf{C}_k \sqrt{\mbf{y}_k} - \mbf{A}_k^{*} (\mbf{I} \otimes \mbf{U}_k^{*}) \text{vec}(\mbf{E}^{*}) \right\rVert^2
\end{align}
with CGLS. For the core tensor $\underline{\mbf{G}}$, note that the function in equation (\ref{eq:tensor_x_k}) is already written in terms of $\text{vec}(\underline{\mbf{G}})$. Hence, the core tensor $\underline{\mbf{G}}$ can be computed by minimizing that function. Lastly, each row of the factor matrix $\mbf{F}$ is updated by solving an $r$-dimensional phase retrieval problem as stated in Section~\ref{sec:algorithm}.

\section{Model Correction Step}
For the experiments pertaining to the CDP measurements, the output of the three algorithms went through a ``model correction'' step. We implemented the same model correction step as proposed by Nayer et al.~\cite{prov_lrpr}, which was taking the output of
any low-rank phase retrieval algorithm (e.g. TSPR, AltMinLowRaP) and running a few iterations of any phase retrieval algorithm (e.g. RWF, TWF) to correct any errors of each image frame that may have been induced by imposing the low rank structure. More specifically, recall that in low-rank phase retrieval, we have measurements generated from the model 
\begin{align}
    \mbf{y}_k = |\mbf{A}_k^{*} \mbf{x}_k|, \, k=1,\ldots, q.
\end{align}
Suppose that with these measurements, we ran TSPR for $T$ iterations, obtaining an output $\underline{\mbf{X}}^T$, where
\begin{align}
    \underline{\mbf{X}}^T = [\mbf{X}_1^T, \mbf{X}_2^T, \ldots, \mbf{X}_q^T].
\end{align}
We can correct any errors of each image $\mbf{X}_k$ by initializing and running any standard phase retrieval algorithm with $\mbf{X}_k^T$ (and with sampling matrix $\mbf{A}_k$ and measurements $\mbf{y}_k$).
One can think of this step as each output frame $\mbf{X}_k^T$ being a ``warm start'' for standard phase retrieval.

Our empirical results showed that this model correction step only worked for the over-determined setting. The reason for this is that since the best sample complexity for phase retrieval is $m \geq Cn$ for some constant $C$, having a warm start would not benefit phase retrieval for the under-determined case. That is, one cannot simply obtain a ``good enough'' output from, for example AltMinLowRaP, and run this correction step in the under-sampled regime.




%



\clearpage
\bibliographystyle{IEEEtran}
\bibliography{refs}

\end{document}